\documentclass[12pt]{iopart}

\usepackage{iopams}

\usepackage{bm}
\usepackage{graphicx}
\usepackage{color}

\newcommand{\blue}[1]{\textcolor[rgb]{0.00,0.00,0.00}{#1}}

\begin{document}

\title[Ionization injection subject to a magnetic field]{Ionization injection in a laser wakefield accelerator subject to a transverse magnetic field}

\author{Q. Zhao$^{1,2}$, S. M. Weng$^{1,2,\dagger}$, Z. M. Sheng$^{1,2,3,6,\ddagger}$, M. Chen$^{1,2}$, G. B. Zhang$^{1,4}$,W. B. Mori$^{5}$, B. Hidding$^{3}$,D. A. Jaroszynski$^{3}$, J. Zhang$^{1,2}$}

\address{$^1$ Key Laboratory for Laser Plasmas (MoE), School of Physics and Astronomy, Shanghai Jiao Tong University, Shanghai 200240, China}%
\address{$^2$ Collaborative Innovation Center of IFSA, Shanghai Jiao Tong University, Shanghai 200240, China}%
\address{$^3$ SUPA, Department of Physics, University of Strathclyde, Glasgow G4 0NG, UK}
\address{$^4$ College of Liberal Arts and Sciences, National University of Defense Technology, Changsha 410073, China}
\address{$^5$ Department of Physics and Astronomy, University of California, Los Angeles, California 90095, USA}
\address{$^6$ Tsung-Dao Lee Institute, Shanghai Jiao Tong University, Shanghai 200240, China}

\ead{\mailto{$^\dagger$wengsuming@gmail.com}, or \mailto{$^\ddagger$z.sheng@strath.ac.uk}}%

\begin{abstract}
The effect of an external transverse magnetic field on ionization injection of electrons in a laser wakefield accelerator (LWFA) is investigated by theoretical analysis and particle-in-cell simulations.
On application of a few tens of Tesla magnetic field, both the electron trapping condition and the wakefield structure changes significantly such that injection occurs over a shorter distance and at an enhanced rate.
Furthermore, beam loading is compensated for, as a result of the intrinsic trapezoidal-shaped longitudinal charge density profile of injected electrons.
The nonlinear ionization injection and consequent compensation of beam loading lead to a reduction in the energy spread and an enhancement of both the charge and final peak energy of the electron beam from a LWFA immersed in the magnetic field.
\end{abstract}


\submitto{\NJP}

\maketitle

\tableofcontents

\section{Introduction}

The laser wakefield accelerator (LWFA) \cite{Tajima1979,RMP81.1229} has attracted growing attention over the last decades because it can sustain ultra-high acceleration gradients (GV/m). The plasma wave in the LWFA is excited by the ponderomotive force of an intense, ultra-short duration  laser pulse. Its phase velocity, $v_\phi$, is close to the group velocity of the laser pulse in plasma. This sets a threshold requirement for injection of electrons into the wakefield;  to be continuously accelerated, they have to move in phase with the wakefield.
An outstanding issue of the LWFA is how to control the injection process, while optimizing the quality of the electron beam produced. In addition to the usual self-injection in the blow-out regime \cite{APB92.075003, Osterhoff}, injection can also be controlled using additional laser pulses \cite{PRL79.2682,Nat444.737}, plasma density transitions \cite{PRE58.5257,PRL86.1011,PRL110.135002,TooleyPRL2017}, external magnetic fields \cite{Hosokai, Froula2009, PRL106.225001}, etc. Controlled ionization injection has also recently been proposed \cite{PRL76.2073,JAP99.056109,PRL104.025003,PRL104.025004,POP19.033101}.

Ionization injection requires inner-shell electrons of high-Z atoms to be released at a particular phase inside the wake bubble.
These electrons will be trapped if they experience sufficiently large  potential difference as they slip backwards to catch up with the wake.
\blue{Since injected electrons are released inside the wake bubble in ionization injection, these electrons can get an additional energy gain due to the potential difference between the edge and the interior of the bubble.
As a results, ionization injection can occur at relatively lower laser intensity and/or lower plasma density in comparison with self-injection \cite{PRL76.2073,JAP99.056109,PRL104.025003,PRL104.025004,POP19.033101}.}
However, ionization injection often results in a large energy spread because electrons are continuously injected at various phases into the wake and experience different accelerating times.
Many schemes have been proposed to reduce the injection distance and the energy spread, such as
using two gas cells to separate the injection and acceleration stages \cite{PRL107.035001,PRL107.045001}, dual-colour lasers to control injection \cite{ZengPRL2015}, or an unmatched laser pulse to truncate the injection \cite{POP21.030701,Srep5.14659}.
However, the narrow energy spread in these schemes is usually achieved at the expense of a lower beam charge.

In this paper, we investigate the effect of an ETMF on ionization injection in the LWFA.
\blue{
In the self-injection scenario, it was previously found that the (longitudinal) trapping condition can be effectively relaxed by an ETMF of a few hundreds of Tesla  \cite{PRL106.225001}.
In comparison, here we find that the ETMF required for tuning the LWFA electron beam can be significantly reduced in the ionization injection scenario.
It is found that a nonlinear ionization injection process, characterized by an enhanced injection rate over a shortened distance, can occur under an ETMF of a few tens of Tesla.
The reduction in the required ETMF is attributed to the reduced self-generated magnetic field in ionization injection, which usually uses lower laser intensity and lower plasma density \cite{PRL76.2073,JAP99.056109,PRL104.025003,PRL104.025004,POP19.033101}.}
More importantly, this nonlinear injection process optimizes the longitudinal beam current profile of the injected electrons, which results in a linearly modified wakefield as a result of beam loading.
Such a linearly modified wakefield effectively suppresses the energy spread because of phase rotation \cite{PRL101.145002,PRL103.194804} and results in dark-current-free bunch generation. Finally, the boosted injection rate, together with the tailored beam loading, allows for simultaneous reduction in the energy spread and enhancement of the beam charge.

\section{Theoretical analysis}
We start by considering the (longitudinal) trapping condition of electrons in the presence of an ETMF. In the frame co-moving with the wake ($x$, $y$, $\xi=z-v_\phi t$), the electron motion is governed by a conservative Hamiltonian $H=\gamma- v_\phi u_z - \psi$ \cite{pop4.217}, where $\gamma = \sqrt{1+u_{\perp}^2+u_z^2}$ is the electron Lorentz factor with the transverse  ($u_{\perp}$) and longitudinal ($u_z$) momenta, $\psi = e(\Phi-v_\phi A_z)$ is the wake potential normalized to $m_ec^2$, and $\Phi$ and $A_z$ are respectively the scalar and vector potentials of the wakefield.
Unless otherwise noted, we use dimensionless units for the equations and variables in the following.
Time, length, velocities, momenta, and magnetic fields, respectively, are normalized to, $1/\omega_p$, $c/\omega_p$, $c$ and $m_ec$, and $m_e\omega_p/e$ with the plasma frequency $\omega_p=(n_0 e^2/\varepsilon_0m_e)^{1/2}$, and the electron density $n_0$, mass $m_e$, and charge $e$.
An electron can be trapped only if its \blue{longitudinal} velocity $v_z$ reaches $v_\phi$ before it slips backwards to the potential through $\psi_{\min} $ \cite{PRL104.025003}.
If there is no ETMF, the \blue{longitudinal} trapping condition can be written as $\Delta \psi=(\psi_{\min}-\psi_i)\leq (1+u_\perp^2)^{1/2}/\gamma_\phi-1$ \cite{PRL98.084801}, where $\psi_i$ is the wake potential at the ionization position and $\gamma_\phi=(1-v_\phi^2)^{-1/2}$.
However, an additional vector potential that satisfies $\nabla \times \bm A^\texttt{ext}=\bm B^\texttt{ext}$ has to be considered in the presence of an ETMF $\bm B^\texttt{ext}$.
Assuming a uniform $\bm B^\texttt{ext}=b_0\bm{\hat{y}}$, the modified \blue{longitudinal} trapping condition is\cite{PRL106.225001}
\begin{equation}
\Delta\psi\leq\frac{\sqrt{1+u_\perp^2}}{\gamma_\phi}-1 + \Delta \Psi^{\rm{ext}}, \label{trapext}
\end{equation}
where $\Delta\Psi^{\rm{ext}}=b_0 v_\phi(x_i-x_f)/2$ is the vector potential difference due to the ETMF, $x_i$ and $x_f$ are the initial ionization and final injection transverse displacements, respectively.
Physically, the ETMF enhances or suppresses electron injection depending on the direction of the longitudinal Lorentz force on the newly-born electrons. Therefore, the modified \blue{longitudinal} trapping condition is relaxed if $x_i>x_f$, and is tightened if $x_i<x_f$.

The longitudinal trapping condition (\ref{trapext}) is a necessary, rather than a sufficient, condition for electron injection. Considering the 3D electron dynamics, injected electrons must also satisfy the transverse trapping condition, i.e., be trapped in the focusing region that is usually located near the bottom of the wake bucket.
\blue{The transverse component of the wakefield can be written as $W_\perp=E_r-B_\theta$ with the radial electric field $E_r$ and the azimuthal magnetic field $B_\theta$ \cite{Katsouleas1987}. In the LWFA, the total wakefield is the superposition of the laser wakefield and the beam wakefield, i.e., the beam loading wakefield. In the focusing region located at the bottom of the wake bucket, $E_r$ is defocusing and $B_\theta$ is focusing, and the total transverse wakefield is focusing since $B_\theta$ is dominant in this region.}
For the sake of simplicity, we ignore the radial electric fields and only consider the magnetic fields that include the self-generated magnetic field $\bm B^\texttt{self}$ and the ETMF $\bm B^{\texttt{ext}}$.
Further, we assume that $\bm B^{\texttt{ext}}= b_0 \bm{\hat{y}}$ as before, while the azimuthal $\bm B^{\texttt{self}}=-br/R_m\bm{\hat\theta}$ has a linear profile until a cut-off radius $R_m$ \cite{pop11.5256}, where $R_m$ is defined as the position where $\bm B^{\texttt{self}}$ reaches its maximum(minimum) value.
Now we are interested in whether an electron in the $y=0$ plane can be transversely trapped or not.
We therefore assume that this electron has already satisfied the longitudinal trapping condition (\ref{trapext}), i.e., $v_z \simeq v_\phi$, at $x=x_\phi$ ($x_\phi$ is the off-axis position where $v_z$ reach $v_\phi$) with the instantaneous transverse momentum $u_{x0}$ and longitudinal momentum $u_{z0} \simeq \gamma_\phi v_\phi$.
As the variation in $\xi$ during one betatron oscillation is usually negligible compared with the betatron oscillation amplitude, the longitudinal momentum variation $U_z(x)\equiv u_z-u_{z0}$, due to the betatron oscillation, is given by
\begin{equation}
U_z(x) = \int_{x_\phi}^{x} \frac{du_z}{dt} \frac{dx}{v_x} \simeq \frac{b(x^2-x_\phi^2)}{2R_m} - b_0(x-x_\phi), \label{Ax}
\end{equation}
where $v_x$ is the velocity in the $x$ direction.
Using the approximation $u_x^2+u_z^2\simeq u_0^2\equiv u_{x0}^2+u_{z0}^2$, one can obtain
\begin{equation}
u_x^2=u_0^2-[u_{z0}+U_z(x)]^2 \simeq u_{x0}^2-2 u_{z0}U_z(x).\label{Tmom}
\end{equation}
If an electron is transversely trapped, it should have two turning points $|x_T| \leq R_m$, where $u_x(x_T)=0$. Equivalently, $U_z(x)=U_0$ has two roots in the region $|x| \leq R_m$, where $U_0\equiv u_0-u_{z0}\approx u^2_{x0}/2u_{z0}$.
This prescribes the following transverse trapping condition
\begin{equation}
b\geq b_{\rm{crit}}= 2R_m( U_0 + b_0 R_m -b_0 x_\phi)/(R_m^2-x_\phi^2),  \label{bmin}
\end{equation}
where $b_{\rm{crit}}$ is the critical amplitude of self-generated magnetic field required for the transverse trapping.
\blue{Since $U_0$ is usually negligible for the injected electrons, $b_{\rm{crit}}$ is roughly proportional to the ETMF amplitude $b_0$. Specially, $b_{\rm{crit}}\simeq2b_0$ for $x_\phi=0$.}

\blue{
In order to inject the electrons into the wake under an ETMF, both the longitudinal trapping condition (\ref{trapext}) and the transverse trapping condition (\ref{bmin}) should be satisfied. On the one hand, the longitudinal trapping condition is relaxed by the ETMF since it contributes an additional vector potential difference \cite{PRL106.225001}. On the other hand, the transverse trapping condition becomes tougher under the ETMF that tends to deflect the electrons, and then a stronger self-generated magnetic field is required to focus the injected electrons.}

\section{PIC simulations}

Three-dimensional PIC-simulations with OSIRIS \cite{Fonseca} have been carried out to visualize the ionization injection under an ETMF.
In each simulation,  \blue{a simulation box with a size of $32.5\times32.5\times12.5$ $(c/\omega_p)^3$ moves along the z-axis at the speed of light, and it is divided into $260\times 260\times 1600$ cells with $1\times1\times 2$ particles per cell, the size of each cell is $0.125\times0.125\times0.0078125$ $(c/\omega_p)^3$}.
We assume a typical laser pulse with parameters of 100 TW and 30 fs incident along the z-axis into a region containing helium-nitrogen mixed gas.
The background helium plasma density is $n_e=1.745\times10^{18}$ $\rm{cm^{-3}}$,
which is doped with 2\% nitrogen atoms.
The laser pulse is linearly polarized and has a wavelength of 0.8 $\rm{\mu m}$ and \blue{spot size} of 30 $\rm{\mu m}$.
The laser power is well above the threshold for relativistic self-focusing (17 TW), and its normalized vector potential $a_0\equiv |eE_0/m_e \omega c|\simeq 1.8$ is close to the ionization threshold of nitrogen inner-shell electrons  \cite{PRL104.025003}.
The plasma is exposed to a uniform ETMF $B_y^{\rm{ext}}$ along the $+\hat{\bm y}$ direction.
We compare the results with $B_y^{\rm{ext}}$=0, 10, 20, and 50 T (corresponding to $b_0\equiv eB/m_e\omega_p \simeq$ 0, 0.024, 0.048, and 0.117).
It is worth pointing out that the applied ETMFs have nearly no impact upon the background plasma since $b_0 \ll 1$, while they may significantly affect the dynamics of ionization injected electrons.

\subsection{Ionization injection under an external magnetic field}

\begin{figure}
\centering\includegraphics[width=0.8 \textwidth]{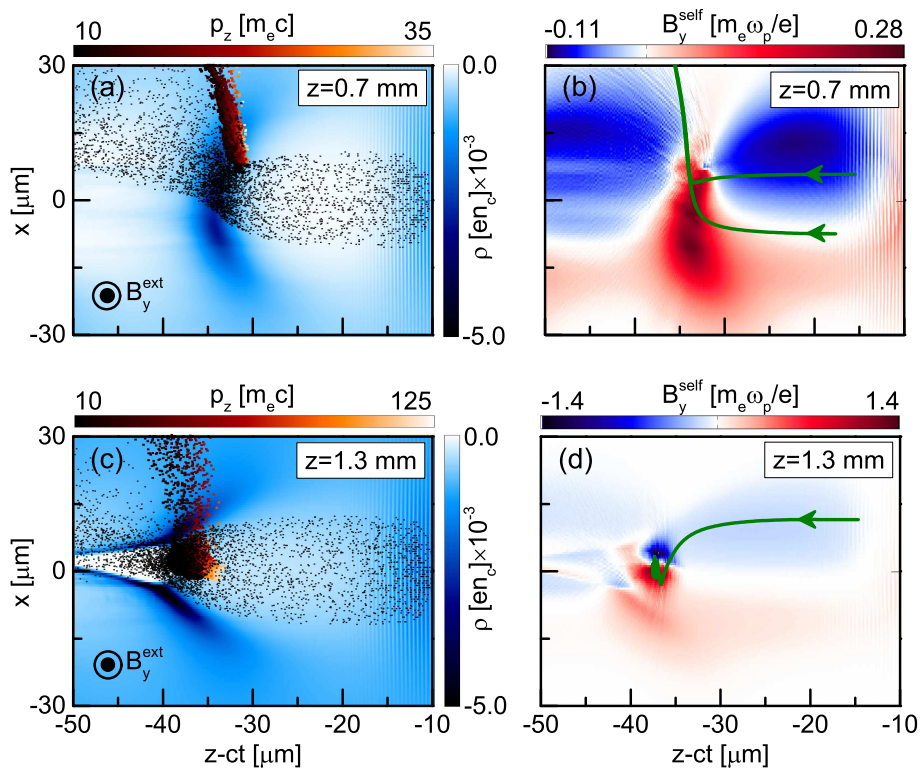}
\caption{(Color online)
\blue{(a) and (c) The distributions of the electron density (blue) in the $y=0$ plane at different propagation distances, superimposed with energetic electrons (color dots) with $\gamma>10$ and low energy electrons (black dots) with $\gamma<10$.
(b) and (d) The distributions of the wakefield azimuthal magnetic field (normalized to $m_e\omega_p/e=4.237\times10^{2}$ T) in the $y=0$ plane at different propagation distances, the olive curves indicate the typical orbits of energetic electrons in the co-moving frame ($x, y, z-ct$). The imposed ETMF is $B_y^{\rm{ext}}=50$ T.}
} \label{FigPhys1}
\end{figure}

Figure \ref{FigPhys1}(a) illustrates that at a propagation distance $z\simeq 0.7$ mm a considerable number of \blue{energetic} electrons have already achieved the wake phase velocity when they slip backwards to the focusing region.
However, these energetic electrons are deflected upwards by the magnetic field and cannot be injected as shown in figure \ref{FigPhys1}(b).
Figure \ref{FigPhys1}(b) also shows that the self-generated magnetic field $B_{y}^{\rm{self}}$ is highly asymmetric about the x-axis at this moment due to the deflection of electrons.
As the laser intensity increases during the self-focusing, \blue{the self-generated magnetic field will increase gradually and
trigger the electron injection as long as it exceeds the critical amplitude required for the transverse trapping.}
The current of injected electrons will enhance the self-generated magnetic field in turn.
\blue{Finally, an avalanche of electron injection occurs when the increasing self-generated magnetic field overwhelms the ETMF.}
Therefore, a large amounts of electrons are successfully injected at $z\simeq 1.3$ mm as shown in figure \ref{FigPhys1}(c), where the bottom of wake bucket is even split apart by the strong Coulomb repulsion force of the injected electrons.

\begin{figure}
\centering\includegraphics[width=0.8 \textwidth]{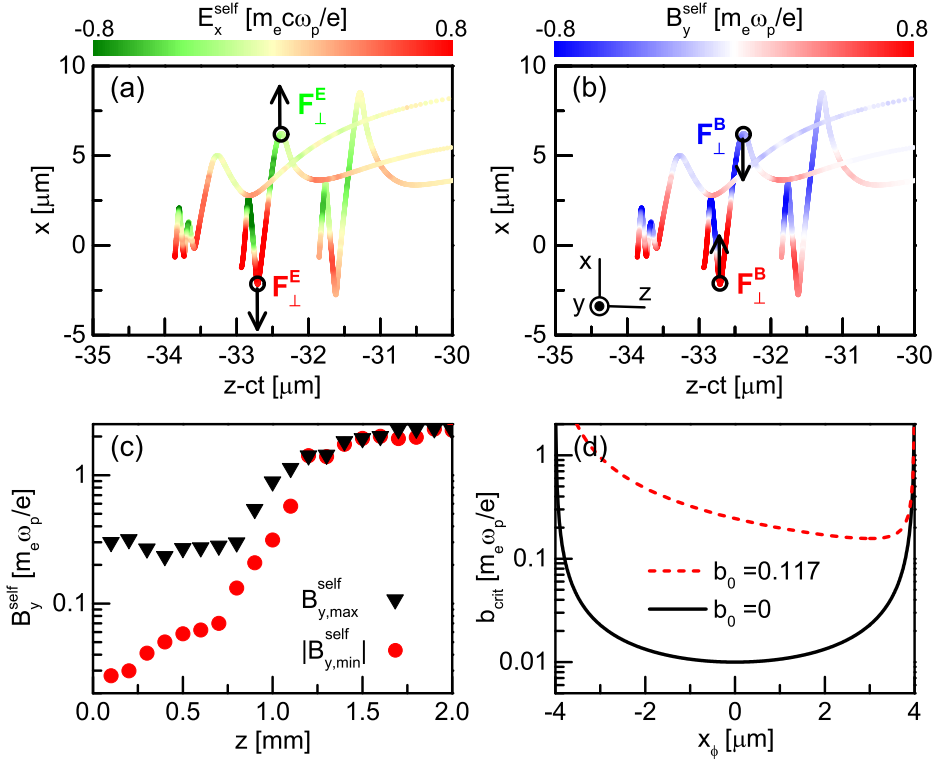}
\caption{(Color online)
\blue{The representative trajectories of two injected electrons in the $y=0$ plane are displayed, with the instantaneous (a) electric field in x direction and (b) magnetic field in y direction.
The black circle denotes the turning points, the arrows denotes the directions of the magnetic force $F_\perp^B=ev_zB_y$ or electric force $F_\perp^E=-eE_x$.}
(c) The minimum negative ($B_{y,\min}^{\rm{self}}$) and maximum positive ($B_{y,\max}^{\rm{self}}$) self-generated magnetic field in the focusing region vs the laser propagation distance.
(d) $b_{\rm{crit}}$ predicted by Eq. (\ref{bmin}) in the cases $b_0$=0 and 0.117 ($B_y^{\rm{ext}}=50$ T).
} \label{FigPhys2}
\end{figure}

\blue{To illustrate that the transverse trapping force of injected electrons is provided by the magnetic force rather than the electric force, the representative trajectories of two injected electrons are displayed in figure. \ref{FigPhys2} (a) and (b) with the instantaneous radial electric field and azimuthal magnetic field, respectively.
It is clear that the electric force $-eE_x$ is defocusing while the magnetic force $ev_zB_y$ is focusing at every turning points of the trajectories.
Therefore, the self-generated azimuthal magnetic field is dominant in the transverse trapping of electrons.}

\blue{To quantitatively analyses the transverse trapping process of electrons, the time evolution of the self-generated magnetic field amplitude is shown in figure \ref{FigPhys2}(c).}
A significant enhancement in $B_y^{\rm{self}}$ due to the electron injection is clearly observed after $z\simeq 0.8$ mm in figure \ref{FigPhys2}(c).
Further, we find from the simulations that $U_0\approx u^2_{x0}/2u_{z0} \simeq 0.02$ and $R_m=4$ $\mu$m are good approximations for the transverse trapping model given above. Substituting these values into Eq. (\ref{bmin}), one can estimate the required $b_{\rm{crit}}$ for the transverse trapping condition.
Figure \ref{FigPhys2}(d) shows that the minimum $b_{\rm{crit}}$ is about $0.18$ for the case $b_0=0.117$ ($B_y^{\rm{ext}}=50$ T), which is one order of magnitude higher than that in the case $b_0=0$.
\blue{
We find that $b_{\rm{crit}}\simeq 0.18$ is roughly approximate to the amplitude of the self-generated magnetic field $B_{y,min}^{\rm{self}} \simeq 0.15$ at $z=0.8 mm$ when the injection is triggered in the simulation case with $B_y^{\rm{ext}}=50$ T.}
Moreover, $b_{\rm{crit}}$ for electrons with $x_\phi>0$ is much smaller than that for electrons with $x_\phi<0$ if $b_0 > 0$. That is to say, electrons from the upper half space ($x>0$) are more easily trapped, which makes injection asymmetric under an ETMF.

\begin{figure}
\centering\includegraphics[width=0.8 \textwidth]{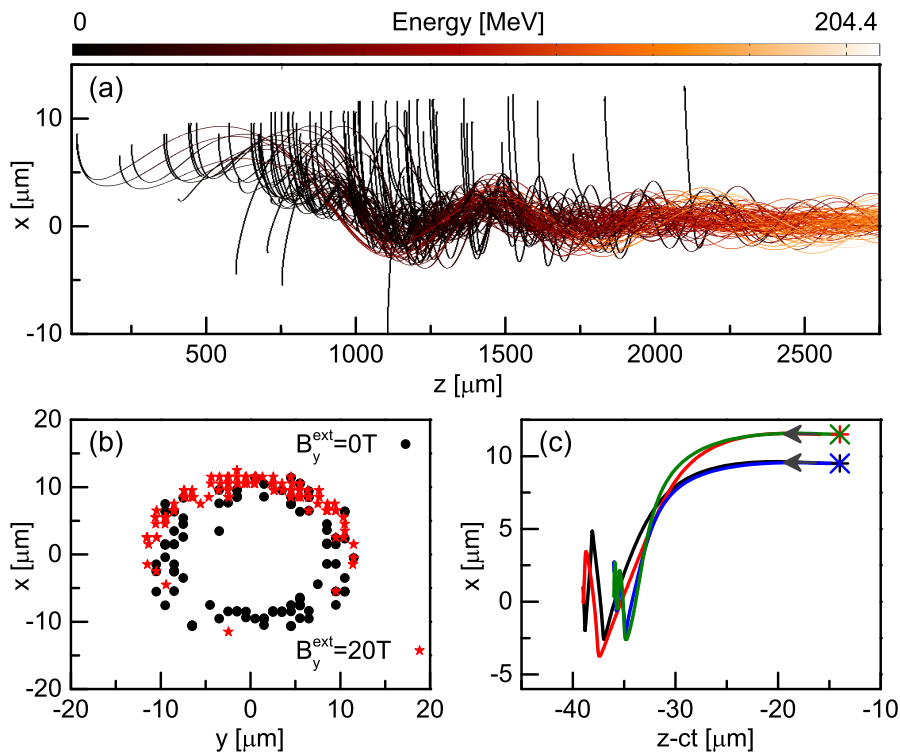}
\caption{(Color online) \blue{(a) The trajectories of randomly selected 100 injected electrons with an ETMF $B_y^{ext}=20T$.} (b) The initial ionization positions of injected electrons in (a) (red stars), in comparison with those without the ETMF (black dots). (c) The trajectories of four typical injected electrons in (a) in the commoving frame. These four electrons originate from two ionization phases, respectively.
 }\label{Figtrack}
\end{figure}

\blue{
The asymmetric injection under an ETMF is illustrated by the trajectories of injected electrons in figure \ref{Figtrack}(a), which shows that the most of injected electrons originate from the upper half space ($x>0$).
It is also seen that the trapped electrons are injected off-axis and oscillate violently before $z\approx 1.5 $ mm since the self-generated magnetic field in the focusing region is highly asymmetric at the early stage.
With the relativistic self-focusing of the laser pulse and the enhancement of the injected electron current, the self-generated magnetic field increases quickly and becomes symmetric. Consequently, the transverse oscillation of electrons will be weakened.}
Figure \ref{Figtrack}(b) compares the initial ionization positions of injected electrons in the cases with and without an ETMF. Without the ETMF, it seems that the injected electrons come from a hollow ring that is roughly symmetric around the laser axis. The electrons ionized near the laser axis are not injected because they do not reach the focusing region due to their small injection positions $|\xi|$.
With the ETMF, however, the injected electrons mainly come from the upper half of the hollow ring due to the asymmetric transverse trapping condition Eq.(\ref{bmin}).

More importantly, the trajectories of injected electrons are more chaotic under the ETMF.
Figure \ref{Figtrack}(c) displays the typical trajectories of four injected electrons from two different ionization phases. It is illuminated that under the ETMF the electrons with the same ionization phase can have completely different longitudinal injection positions, which is distinct from the case without the ETMF.
This is because the self-generated magnetic field in the focusing region rapidly increases and is highly asymmetric under an ETMF as shown in figure \ref{FigPhys2}.

\subsection{Nonlinear injection rate and modified charge profile}

\begin{figure}
\centering\includegraphics[width=0.8 \textwidth]{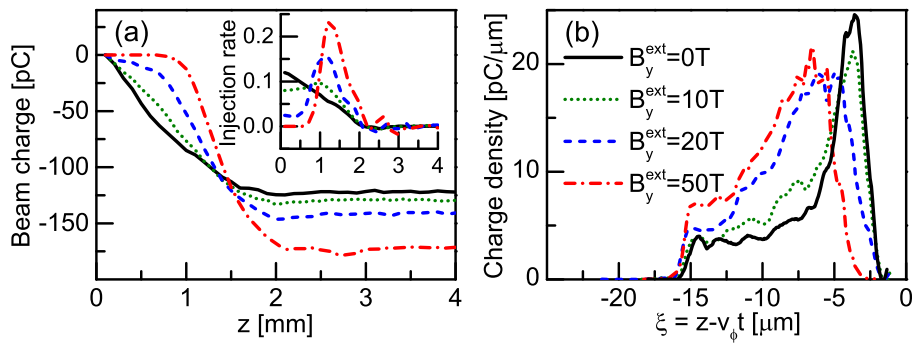}
\caption{(Color online) (a) The beam charge vs the laser propagation distance under $B_y^{\rm{ext}}=$0, 10, 20 and 50 T. Inset: The corresponding injection rates in units of nC/mm. (b) The charge density profiles of loaded electron beams.}\label{FigRate}
\end{figure}

The above analysis illuminates that the electron injection becomes efficient only if the self-generated magnetic field $B_y^{\rm{self}}$ overwhelms the ETMF $B_y^{\rm{ext}}$, which tends to deflect the electrons.
Therefore, at early stages the injection rate decreases with the increasing $B_y^{\rm{ext}}$, and virtually no injection occurs when $B_y^{\rm{ext}}=50 $ T in figure \ref{FigRate}(a).
Due to relativistic self-focusing, however, injection will be triggered at $z \sim 0.8$ mm when $|B_{y,\min}^{\rm{self}}|$ in figure \ref{FigPhys2}(c) is comparable to the minimum $b_{\rm{crit}}$ when $B_y^{\rm{ext}}=50 $ T in Figs. \ref{FigPhys2}(d) and injection enhances $B_y^{\rm{self}}$ in return. In this case, nonlinear injection occurs because of the increasing injection rate, which is evident in the inset of figure \ref{FigRate}(a).
Furthermore, the longitudinal trapping condition (\ref{trapext}) can be relaxed because most of the injected electrons under an ETMF come from the upper half space and satisfy $x_i>x_f$, which results in the peak of the injection rate being enhanced.
In all cases, the injection rates decrease in the latter stages because the beam loading effect undermines the accelerating field \cite{POP19.033101}. Note that the total charge for an ETMF of 50 T can reach 175 pC.

Not only can the ETMF shorten the injection distance without reducing beam  charge, but it also shapes the beam density profile ideally for high beam quality and acceleration efficiency.
Without the ETMF, the relative longitudinal injection positions $\xi$ of electrons in the blown-out regime can be determined by \cite{PRL117.034801}
\begin{equation}
\xi=-\sqrt{4+\xi_i^2+r_i^2-r^2-4(\gamma-v_\phi u_z)}, \label{xiEq}
\end{equation}
where $\xi_i$ ($r_i$) and $\xi$ ($r$) are respectively the longitudinal (transverse) coordinates when the electrons are initially ionized and finally injected, and the term $\gamma-v_\phi u_z$ is negligible for the electrons that have just been loaded.
Without the ETMF, a lot of ionized electrons can be easily injected within a propagation distance as short as a few hundreds of micrometers due to the looser transverse trapping condition shown in figure \ref{FigPhys2}(d).
According to Eq. (\ref{xiEq}), these injected electrons will be loaded at the beam front with relatively small longitudinal coordinates $|\xi|$ since they are ionized at an early stage with relatively small radii $r_i$.
In contrast, under the ETMF most ionized electrons can only be injected after a propagation distance as large as one millimetre. So they are usually ionized at relatively large radii due to the enhanced laser intensity by the self-focusing.
Figure \ref{Figtrack}(b) demonstrates that the mean ionization radius of injected electrons under the ETMF ($\sim 11.04\mu m$) is a little larger than that without the ETMF ($\sim 9.84\mu m$). Following this larger mean ionization radius, these electrons will be loaded at more lagged phases, according to Eq. (\ref{xiEq}).
Furthermore, these electrons will be distributed into a relatively broad range of longitudinal coordinates due to the uncertain relation between their ionization and injection positions under the ETMF, as shown in figure \ref{Figtrack}(c).
That is to say, the longitudinal charge profile of loaded electron beams can be modified to some extent by an ETMF.
Figure \ref{FigRate}(b) compares the charge density profiles of injected electron beams under different ETMFs.
Without the ETMF, the injected electrons will form a sharp peak at the beam front due to their relatively small ionization radius.
In contrast, trapezoidal-like charge profiles can be formed by the injected electrons in the cases with appropriate ETMFs.

\subsection{Correlation between energy spread and charge profile}

\begin{figure}
  \centering
  \includegraphics[width=0.6 \textwidth]{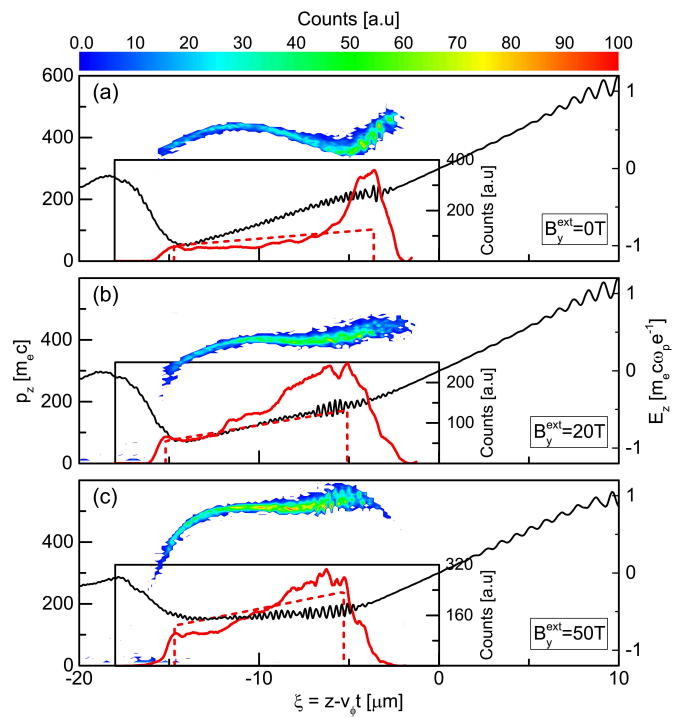}
\caption{(Color online) (a) - (c) Phase-space of injected electrons (color contour) and the accelerating field $E_z$ (black curve) for $B_y^{\rm{ext}}$=0, 20, 50 T. Insets compare the charge profiles of electron beams from the simulations (red solid curve) with the optimized trapezoidal-shaped profiles (red dash curve) predicted by Eq. (\ref{Trapezoid}), where $R_b$ and $r_t$ are obtained from the simulations.}\label{FigPhase}
\end{figure}

The charge profile of loaded electrons  can have a significant effect on the accelerating efficiency and beam quality, because of modifications of the wakefield \cite{PRL101.145002,PRL112.035003,Manahan2016}.
In the ionization injection regime, the energy spread of LWFA electrons arises from two causes \cite{POP19.033101}.
The first is due to the different accelerating times for electrons that are ionized and therefore injected at adjacent phases; while the second cause is due to the different accelerating fields for electrons that are ionized and injected at various phases.

In figure \ref{FigPhase}, we display the distributions of injected electrons in the $\xi-p_z$ phase space for different ETMFs.
Without the ETMF, the energy spread is mainly due to the first cause because most of the electrons are injected within a narrow $\xi$ interval at the beam front. These electrons are injected at the various moments and experience different accelerating times. As a consequence, they will have a broad range of momenta and form a steep slope in the $\xi-p_z$ phase space at the beam front.

\begin{figure}
  \centering
  \includegraphics[width=0.8 \textwidth]{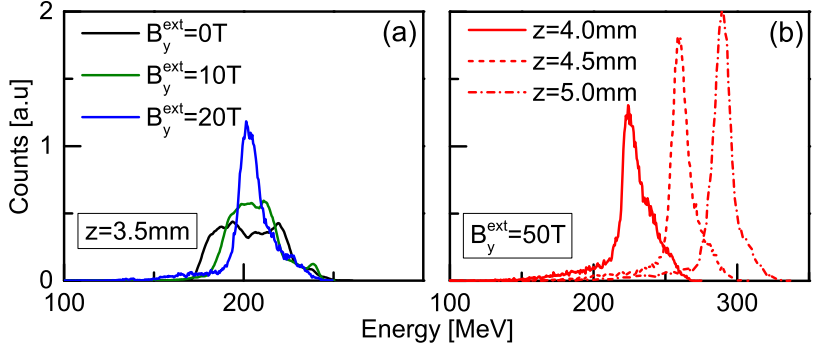}
\caption{(Color online) Energy spectra of electron beams (a) with different ETMF for the same propagation distance, (b) with $B_y^{\rm{ext}}$=50 T for different propagation distances.}\label{FigSpectra}
\end{figure}

In contrast, with a strong ETMF, electrons are loaded into a relatively broad range of $\xi$. Therefore, the energy spread in this case is mainly due to the second cause.
On the other hand, the wakefield can be optimized by the modified charge profile of loaded electrons under a strong ETMF.
In the case of $B_y^{\rm{ext}}=20$ T, it is found that the strength of the wakefield within the electron beam increases linearly with the phase lag $|\xi|$. Taking advantage of such a linearly modified wakefield, the energy spread of electrons can be greatly reduced after they are injected.
In figure \ref{FigSpectra}(a), we compare the energy spectra of electron beams under different ETMFs.
It is illustrated that the energy spread at the propagation distance $z\simeq 3.5$ mm decreases with the increasing ETMF, and a quasi-monoenergetic electron beam can be achieved under an ETMF $B_y^{\rm{ext}}=20$ T.
If the slope of the linearly modified wakefield is too large, however, longitudinal phase mixing will occur due to the strong rotation of loaded electrons in phase space. This kind of phase mixing will increase the energy spread  at a later stage.
Fortunately, under a stronger ETMF $B_y^{\rm{ext}}=50$ T a nearly uniform wakefield $E_z$ is presented within the electron beam in figure \ref{FigPhase}(c).
Theoretically, such a uniform wakefield is achieved by a trapezoidal-shaped beam charge profile \cite{PRL101.145002}
\begin{equation}
\lambda(\xi)=(R_b^4+r_t^4)/8r_t^2-\sqrt{(R_b^4-r_t^4)/8r_t^2}(\xi_t-\xi), \label{Trapezoid}
\end{equation}
where $R_b$ is the radius of the blow-out region, and $r_t=r_b(\xi_t)$ is the channel radius at $\xi_t$ where the loading starts.
The charge profiles from the simulations without and with an ETMF are compared with optimized trapezoidal-shaped charge profiles in the insets of Figs. \ref{FigPhase}(a)- (c), respectively.
It is confirmed that the charge profile for $B_y^{\rm{ext}}$=50 T is in rough agreement with the prediction by Eq. (\ref{Trapezoid}), which is of great benefit to the accelerating efficiency and beam quality.
As shown in figure \ref{FigSpectra}(b), it is demonstrated that the relative energy spread can gradually decrease from $6.2\%$ ($z=4$ mm) to $4.3\%$ ($z=5$ mm), while the peak energy gradually increases from $\sim 224$ to $\sim 290$ MeV.

\subsection{Magnetic effect on transverse emittance}

\begin{figure}
  \centering
  \includegraphics[width=0.8 \textwidth]{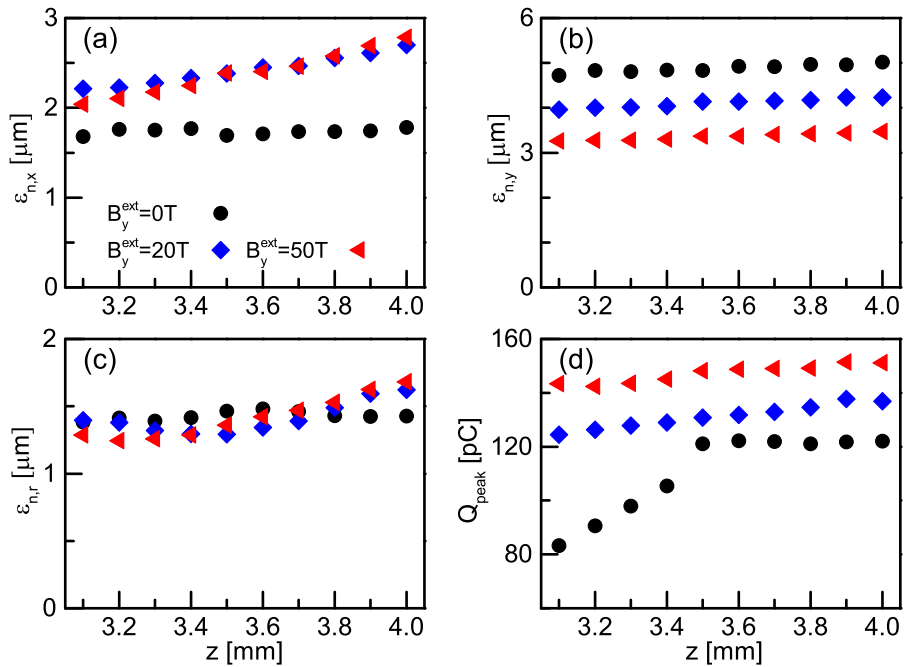}
\caption{(Color online) Time evolution of the normalized RMS transverse emittance of injected electrons of the (qusi)monoenergetic peak in (a) x direction, (b) y direction and (c) radial direction. The transverse beam emittances are defined as $\epsilon_{n,s}=\sqrt{\langle s^2\rangle\langle p_s^2\rangle-\langle sp_s\rangle^2}$, for $s=x$, $y$, and $r$ \cite{Floettmann2003}.(d)Time evolution of charge of the (qusi)monoenergetic peak.}\label{Figemittance}
\end{figure}

As another important property of electron beams, the transverse beam emittances of injected electron beams under different ETMFs are compared in figure \ref{Figemittance}(a)-(c), \blue{ while the electron beam charges are compared in figure \ref{Figemittance}(d). In the calculation of the emittance and the charge, we only consider the electrons in the quasi-monoenergetic part of the beam. Above all, we find that the emittance generally increases with the increasing charge in each case with an ETMF.
In the case without the ETMF, however, the relationship between the emittance and the charge seems vague. This may be because the injected electrons are not so monoenergetic and their distribution in the phase space evolves obviously in this case.
}
Figure \ref{Figemittance}(a) shows that the beam emittance $\epsilon_{n,x}$ in the $x$ direction, that perpendicular to the ETMF direction, will be slightly increased in an ETMF.
\blue{
This may be because the ETMF makes the focusing force nonlinear in this direction by its contribution $ev_zB_y^ext$. As a result, the electron oscillation and the emittance are increased in this direction.}
\blue{
In contrast, figure \ref{Figemittance}(b) shows that the beam emittance $\epsilon_{n,y}$ in the $y$ direction, that parallel to the ETMF direction, will be slightly decreased in an ETMF.
The suppression of $\epsilon_{n,y}$ might be attributed to the stronger self-generated magnetic field under an ETMF.
Finally, figure \ref{Figemittance}(c) indicates that the total transverse beam emittance $\epsilon_{n,r}$ will increase slightly with the increasing ETMF.}


\section{Discussions}
Equation (\ref{bmin}) indicates that the self-generated magnetic field $B^{\rm{self}}$ required for transverse trapping increases with the ETMF $B^{\rm{ext}}$, which implies that the ETMF should not be too strong, otherwise the ionization injection can never be triggered.
Using the laser and plasma parameters of figure \ref{FigPhys1}, we find that no electron injection occurs for $B_y^{\rm{ext}}\geq$ 100 T.
In the self-injection scenario of the LWFA, however, an ETMF of a few hundreds of Tesla is beneficial to electron injection \cite{PRL106.225001}.
\blue{
This is because it is relatively hard to achieve the longitudinal trapping condition in the usual self-injection, and a strong ETMF can greatly relax the longitudinal trapping condition by an additional vector potential difference.
In contrast, the ETMF effect upon the longitudinal trapping condition is not so important in ionization injection since the injected electrons are released inside the wake and they are relatively easier to achieve the phase velocity of the wake in this scenario. As a result, the ETMF mainly appears to modify the transverse trapping condition in the ionization injection.}

\blue{In contrast to self-injection, ionization injection significantly reduces the required ETMF for tuning the LWFA electron beam. In order to dynamically control the transverse trapping condition and then modify the beam quality, we find that the ETMF should be on the order of the self-generated magnetic field according to Eq. (\ref{bmin}).
In the self-injection, the self-generated magnetic field usually is very large since the laser intensity and the plasma density are relatively high in this scenario.
In contrast, the self-generated magnetic field in the ionization injection is relatively small since a lower laser intensity and/or a lower plasma density could be employed in this scenario.
As a result, an ETMF on the order of a few tens Tesla is enough to modify the beam quality in ionization injection.
At the early stage, the electron injection can be effectively suppressed by such an ETMF.
Due to relativistic self-focusing, the injection rate will be dramatically increased as long as the increasing self-generated magnetic field is comparable to the ETMF.}

It is worth pointing out that the strong ETMF offers a new freedom to control ionization injection in a LWFA.
Previously, a few novel schemes have already been proposed to control the ionization injection process and then reduce the energy spread \cite{PRL107.035001,PRL107.045001,ZengPRL2015,POP21.030701,Srep5.14659}.
However, these schemes usually only consider the first cause of energy spread, and narrow the difference in the accelerating time by reducing the injection distance.
However, using an appropriate ETMF, one can not only narrow the difference in the accelerating time via compressing the ionization injection process, but also provide a uniform  accelerating field by optimizing the charge profile of loaded electrons.
These two aspects are the unique advantages of  magnetic controlled ionization injection for a LWFA.

In addition, \blue{we notice that strong magnetic fields on the order of a few tens of Tesla in a small volume can be generated by discharging a high-voltage capacitor through a small wire-wound coil in laboratories \cite{Fiksel2015,Pollock2006,HMFS}}, and a pulsed non-destructive magnetic field above 100 Tesla was recently recorded in the Pulsed Field Facility at Los Alamos National Laboratory \cite{PFF}. Such high magnetic fields are of great interest for controlling laser-plasma interactions \cite{WengOptica,Hu2015}. Particularly, they could provide an alternative powerful means to control the ionization injection and modify the wakefield structure in the LWFA.

\section{Conclusion}
In summary, we have proposed a magnetic-controlled ionization injection scheme for the LWFA.
Under an ETMF, electron trapping occurs only when the self-generated magnetic field is larger than certain critical value as described by equation (\ref{bmin}).
Due to relativistic self-focusing, the increasing self-generated magnetic field triggers electron injection at a particular propagation distance.
As soon as injection is triggered, the current of the injected electrons rapidly enhances the self-generated magnetic field, which in turn, leads to an avalanche of electron injection.
As a result, a large number of electrons are injected over a limited distance.
Moreover, the injected electrons form a trapezoidal-shaped charge profile for appropriate ETMFs.
Such an optimized charge profile can modify the accelerating field to be nearly constant along the propagation direction, which increases the electron energy and, in addition, reduces the energy spread. Consequently, our scheme allows for the generation of high-energy, high-charge beams with narrow energy spread.
\blue{
More importantly, ionization injection in our scheme significantly reduces the ETMF required for tuning the LWFA electron beam in comparison with the self-injection.}

\section*{Acknowledgments}
\addcontentsline{toc}{section}{Acknowledgments}

We thank L. L. Yu, M. Zeng, F. Y. Li, and J. Luo for fruitful discussions.
The work was supported by the National Basic Research Program of China (Grant No. 2013CBA01504), National Natural Science Foundation of China (Grant Nos. 11675108, 11655002, 11721091, and 11774227), National 1000 Youth Talent Project of China, Science and Technology Commission of Shanghai Municipality (Grant No. 16DZ2260200), Science Challenge Project (No.TZ2018005), UK Engineering and Physical Sciences Research Council (EPSRC) (Grant No. EP/N028694/1), EC's H2020 EuPRAXIA (Grant No. 653782) and LASERLAB-EUROPE (Grant No. 654148).
Simulations have been carried out on the Pi supercomputer at Shanghai Jiao Tong University.
The data that support the findings of this study are available at DOI: 10.15129/4404a563-b5d3-490b-b80d-7bf260fbaf59.

\end{document}